\documentclass[showpacs,10pt,twocolumn,prl]{revtex4-1}

\usepackage{amsmath}
\usepackage{amssymb}
\usepackage{graphicx}
\usepackage{graphics}
\usepackage{epsfig}
\usepackage{CJK}
\usepackage{color}
\usepackage{epstopdf}

\begin{document}
\begin{CJK*}{GBK}{}

\title{Emergence of Double-Dome Superconductivity in the Pressurized Dirac Semimetal BaMg$_{2}$Bi$_{2}$}
\author{Qi Wang $^{1,2,\dagger}$, Juefei Wu $^{1,\dagger}$, Cuiying Pei $^{1}$, Yi Zhao $^{1}$, Yiyan Wang $^{3,*}$, and Yanpeng Qi$^{1,2,4,*}$}
\affiliation{
$^{1}$ State Key Laboratory of Quantum Functional Materials, School of Physical Science and Technology, ShanghaiTech University, Shanghai, China\\
$^{2}$ ShanghaiTech Laboratory for Topological Physics, ShanghaiTech University, Shanghai, China\\
$^{3}$ Anhui Provincial Key Laboratory of Magnetic Functional Materials and Devices, Institutes of Physical Science and Information Technology, Anhui University, Hefei, Anhui, China\\
$^{4}$ Shanghai Key Laboratory of High-resolution Electron Microscopy, ShanghaiTech University, Shanghai, China\\
\textbf{Correspondence}: Yanpeng Qi (qiyp@shanghaitech.edu.cn); Yiyan Wang (wyy@ahu.edu.cn)\\
$\dag$ These authors contributed equally to this work.
}

\date{\today}

\begin{abstract}

Dirac semimetal BaMg$_{2}$Bi$_{2}$ is reported to be a unique topological material that manifests surface superconductivity that coexists with bulk band topology at ambient pressure.
Here, we present a comprehensive investigation of high-pressure superconducting properties in BaMg$_{2}$Bi$_{2}$ single crystal. 
Significantly, a pressure-driven double-dome superconducting behavior was revealed, with the superconducting transition temperature $T_{\rm c}$ approaching the maximum values of $\sim$ 6.67 K at 4.5 GPa and $\sim$ 7.22 K at 10.4 GPa for the first and second superconducting domes, respectively. The combination of high-pressure X-ray diffraction, Hall resistivity measurements, and theoretical calculations demonstrates that, the first superconducting regime is closely related to the pressure-modulated Lifshitz transition, whereas the second superconducting phase emerges concurrently with a structural transition from the ambient-pressure $P \bar{3}m1$ phase to a high-pressure $Pnma$ phase.

\end{abstract}

\maketitle

\end{CJK*}

\textbf{Introduction}

Nowadays, the emerging topological semimetals, including Dirac semimetals, Weyl semimetals, and so on, are regarded as the hot topic in the field of condensed matter physics.\cite{021004,303001,015001} Topological electronic structure that manifests band linear dispersion close to the Fermi energy ($E_{\rm F}$) emerges with abundant intriguing quantum states, such as extremely large magnetoresistance, quantum oscillations, large intrinsic anomalous Hall effect, chiral anomaly etc.\cite{280,205105,3681,1125} More notably, the superconductivity that revealed in topological materials has attracted extensive attention, providing a promising platform for realizing the unconventional or topological superconductivity.\cite{076501} 

Generally, the superconducting state identified in topological materials is usually a bulk characteristic, as observed in Weyl semimetal MoTe$_{2}$ with $T_{d}$ phase\cite{11038}, topological Semimetal ZrTe$_{2}$\cite{2301332}, kagome superconductors AV$_{3}$Sb$_{5}$ (A = K, Rb, Cs) etc.\cite{247002, 037403, 034801, 2102813} Recent studies have revealed that superconductivity can also exist in the surface state.
For example, Dirac semimetals KZnBi and SrCuBi were reported to demonstrate surface superconductivity with superconducting transition temperature $T_{\rm c}$ of 0.85 and 2.1 K, respectively.\cite{021065,2400428} The superconductivity was revealed on the surface of topological MoTe$_{2-x}$S$_{x}$ single crystal using scanning tunneling spectroscopy (STM).\cite{115} Weyl semimetal t-PtBi$_{2}$ exhibits a surface superconductivity at 5 K, accompanied by large gap values certified by STM.\cite{9895}

Recently, Dirac semimetals XMg$_{2}$Bi$_{2}$ (X = Ba, Sr), which crystallize in the space group of $P \bar{3}m1$ (No. 164) have elicited widespread attention.\cite{612} 
Structurally, the buckled Mg-Bi honeycomb bilayer and the Ba/Sr layer stack alternatively along the $c$ axis. Angle-resolved Photoemission Spectroscopy (ARPES) experiments combined with the theoretical calculations confirmed the existence of Dirac fermions near $E_{\rm F}$ along the $\Gamma$-$A$ direction in BaMg$_{2}$Bi$_{2}$.\cite{21937} In addition to the topological characteristic in bulk state, a 2D surface superconductivity with $T_{\rm c}$ of approximately 4.77 K was observed in transport measurements.\cite{2208616} It is intriguing to explore the pressure responses of superconductivity and band topology in BaMg$_{2}$Bi$_{2}$ by utilizing the physical pressure without introducing impurity scattering or disorder effects. In this work, we systematically investigated the electrical transport properties in BaMg$_{2}$Bi$_{2}$ single crystal under high pressure. It has been observed that BaMg$_{2}$Bi$_{2}$ exhibits a double-dome superconductivity within the applied pressure. With the combination of the high-pressure X-ray diffraction (XRD), Hall effect measurements, and theoretical analyses, the first superconducting phase, where BaMg$_{2}$Bi$_{2}$ maintains the stability of the $P \bar{3}m1$ crystal structure, is closely related to the Lifshitz transition. Upon further compression, the re-enhancement of superconductivity occurs simultaneously with the structural phase transition.

\begin{figure*}
	\centerline{\includegraphics[scale=0.23]{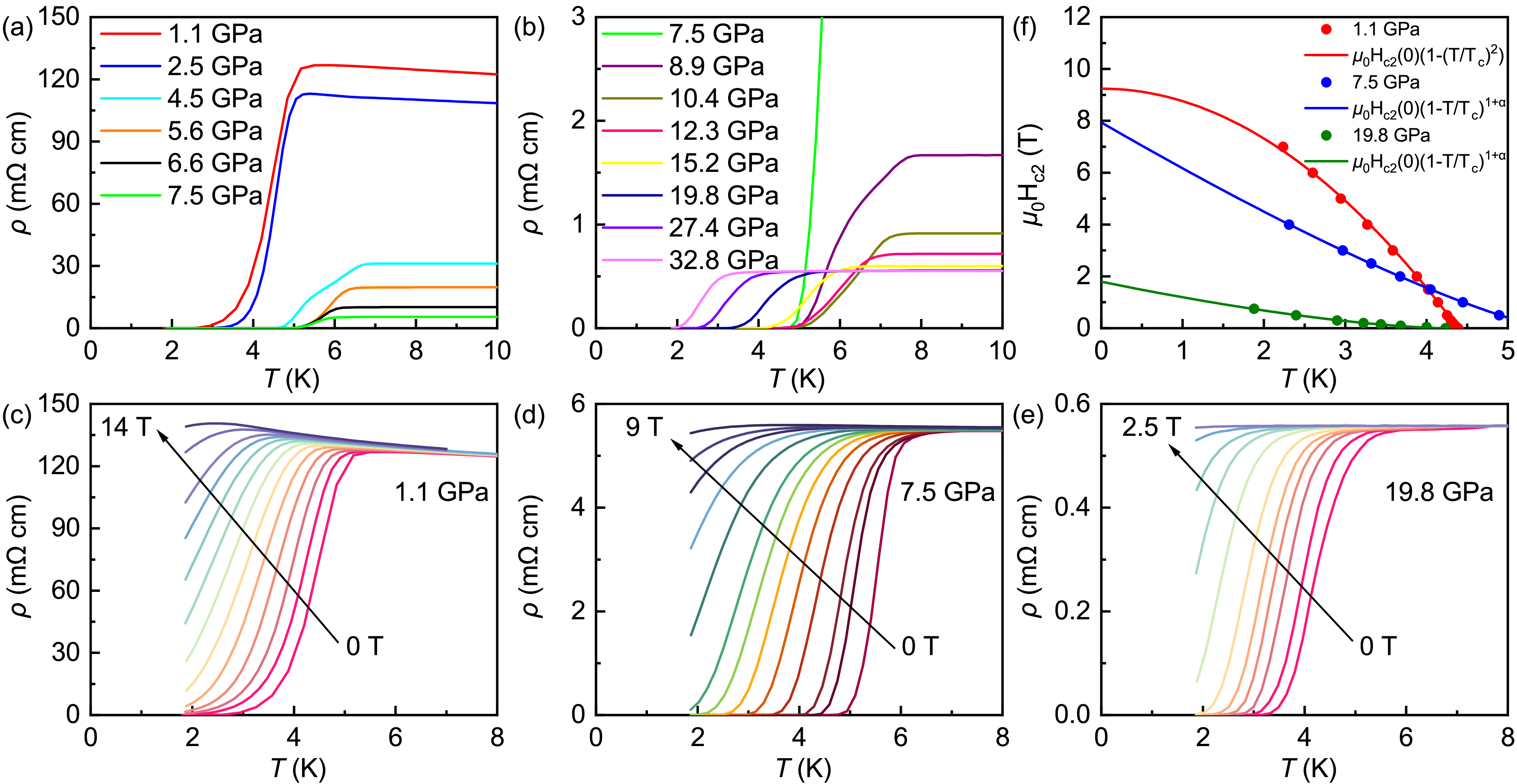}}
	\caption{ (a) and (b) Temperature dependence of resistivity $\rho (T)$ at various pressures in the temperature range from 1.8 to 10 K for BaMg$_{2}$Bi$_{2}$ single crystal. (c)-(e) Temperature dependence of $\rho (T)$ under different magnetic fields at 1.1, 7.5, and 19.8 GPa, respectively. (f) The upper critical field $\mu_{0}$H$_{c2}(T)$ as a function of temperature at representative pressures. The solid lines demonstrate the fitting results using $\mu_{\rm 0} H_{\rm c2}(T) = \mu_{\rm 0} H_{\rm c2}(0)(1-(\frac{T}{T_{\rm c}})^{2})$ and $\mu_{\rm 0} H_{\rm c2}(T) = \mu_{\rm 0} H_{\rm c2}(0)(1-\frac{T}{T_{\rm c}})^{1+\alpha}$, respectively. }
	\label{}
\end{figure*}

\begin{figure*}
	\centerline{\includegraphics[scale=0.25]{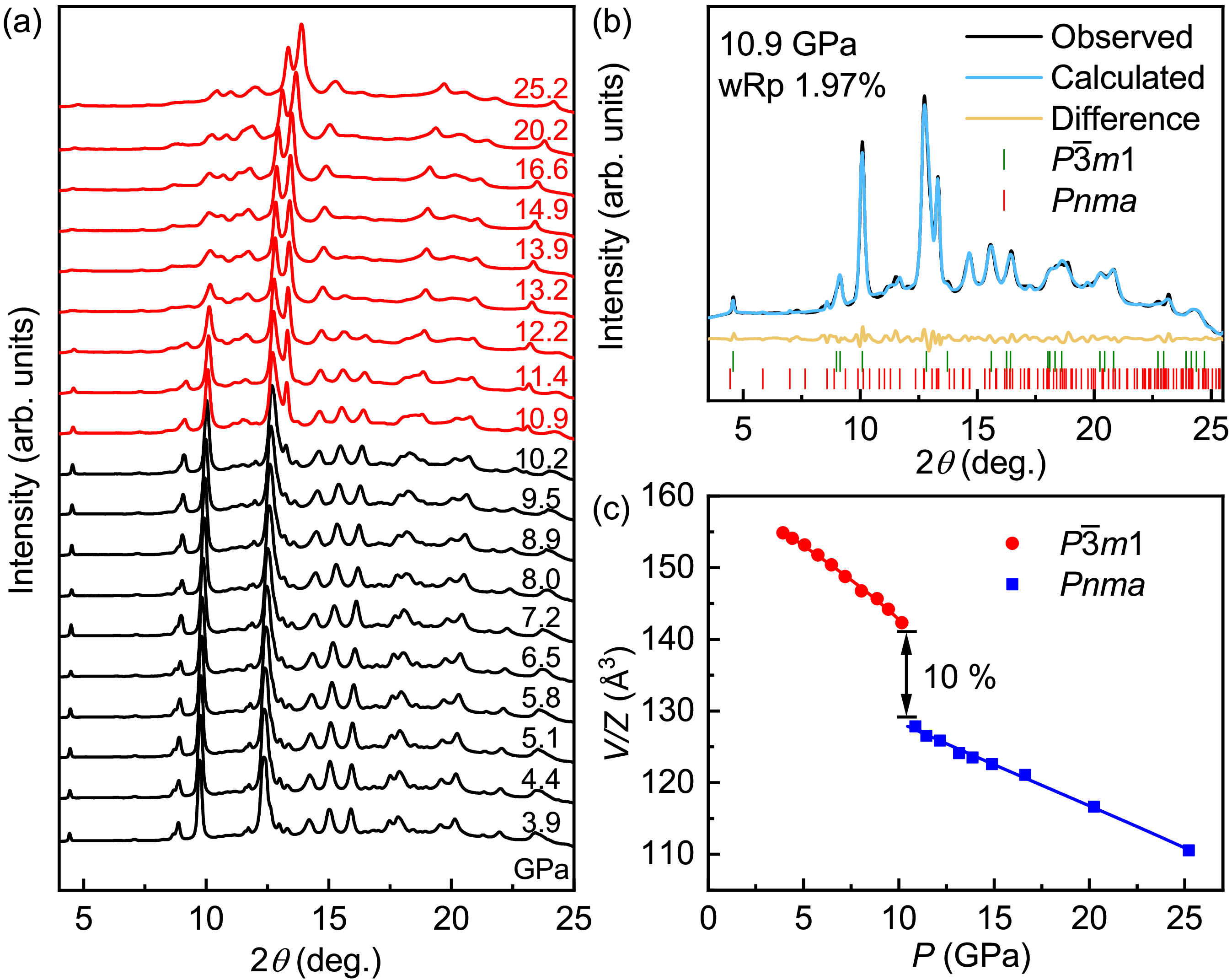}}
	\caption{(a) High-pressure XRD patterns of BaMg$_{2}$Bi$_{2}$ single crystal in the pressure range of 3.9 - 25.2 GPa at room temperature. (b) Rietveld refinement of XRD at 10.9 GPa. (c) The volume per formula $V/Z$ as a function of pressure. The solid lines represent the third-order Birch-Murnaghan fitting.}
	\label{}
\end{figure*}

\textbf{Experimental and Theoretical Methods}

BaMg$_{2}$Bi$_{2}$ single crystals were grown by the Bi flux method.\cite{2208616} The high-pressure resistivity and Hall resistivity measurements were carried out in Quantum Design Physical Property Measurement System (PPMS). The small pieces of BaMg$_{2}$Bi$_{2}$ single crystals were placed in diamond anvil cell (DAC) with 300-$\mu$m diameter anvil culet. The van der Pauw method was used to detect the electrical transport properties under high pressure. High-pressure X-ray diffraction (XRD) measurements were conducted at beamline BL15U of Shanghai Synchrotron Radiation Facility, utilizing an X-ray wavelength of 0.6199 $\mathring{\rm A}$. The BaMg$_{2}$Bi$_{2}$ single crystals were thoroughly ground in the glove box, then the ground powder samples were loaded into the symmetric DAC with 200-$\mu$m diameter anvil culet. Thereinto, the pressure values were calibrated using Ruby luminescence at room temperature.\cite{4673} Due to the air-sensitive characteristics for BaMg$_{2}$Bi$_{2}$ single crystal, all cell preparations were performed in the glove box with the protected atmosphere of argon gas. 

We applied the machine learning graph theory accelerated crystal structure search method (MAGUS) \cite{nwad128,817} to perform the fixed composition structures searches for BaMg$_{2}$Bi$_{2}$ under 20 and 50 GPa, respectively, and over 1000 structures were evolved within 30 generations at each pressure. The MAGUS runs were combined with the Vienna Ab initio Simulation Package (VASP) based on the density functional theory \cite{11169,5188} for structures relaxations. The exchange-correlation functional was the generalized gradient approximation (GGA) under the framework of Perdew, Burkey, and Ernzerhof.\cite{3865} The projector-augmented wave (PAW) approach \cite{17953} described the valence electrons of 5$s^{2}$5$p^{6}$6$s^{2}$ in Ba atoms, 3$s^{2}$ in Mg atoms and in 6$s^{2}$6$p^{3}$ Bi atoms. The maximum cutoff energy and the Brillouin zone sampling were 400 eV and 2$\pi$ $\times$ 0.05 $\mathring{\rm A}^{-1}$ for structure searches, respectively. We recalculate the enthalpy of the BaMg$_{2}$Bi$_{2}$ structures under high pressure by VASP. The plane-wave kinetic-energy cutoff was set to 800 eV and the Brillouin zone sampling was 2$\pi$ $\times$ 0.03 $\mathring{\rm A}^{-1}$. The convergence tolerance for enthalpy calculations was 10$^{-6}$ eV for total energy and 0.003 eV $\mathring{\rm A}^{-1}$ for all forces. We carried out the phonon spectra calculations for the predicted structures using the PHONOPY program package \cite{108}, with the supercell of 2 $\times$ 2 $\times$ 2. We used a denser $k$-mesh grid of 2$\pi$ $\times$ 0.015 $\mathring{\rm A}^{-1}$ for electronic structures calculations, and 2$\pi$ $\times$ 0.01 $\mathring{\rm A}^{-1}$ for Fermi surface calculations. The spin-orbital coupling (SOC) effect was taken into consideration in the electronic structure calculations, and the total energy is further converged to 10$^{-8}$ eV.

\textbf{Results and Discussion}

\textbf{High-Pressure Electrical Transport}

Figure 1a,b illustrate the resistivity $\rho (T)$ as a function of temperature from 1.1 to 32.8 GPa. It is apparent that the superconducting feature of BaMg$_{2}$Bi$_{2}$ single crystal can be effectively modulated by the physical pressure. Initially, $T_{\rm c}$ gradually increases with increasing pressure, the onset superconducting temperature $T_{\rm c,onset}$ approaches a maximum value of 6.67 K around 4.5 GPa and the maximum zero-resistivity temperature $T_{\rm c,zero}$ reaches 4.77 K at 5.6 GPa. Subsequently, it begins to decrease upon further pressurization. Intriguingly,for pressure above about 7.5 GPa, $T_{\rm c}$ shows an abnormal rising behavior and reaches the highest value of 7.22 K at approximately 10.4 GPa. Then, it exhibits a monotonic reduction at higher pressure. Notably, the normal-state $\rho (T)$ curves exhibit semiconducting-like behavior below 2.5 GPa as shown in Figure S1, which has also been observed under ambient pressure \cite{2208616}. When the pressure increases further, the metallic behavior appears in the whole temperature region. In addition, the repeated high-pressure measurements of $\rho (T)$ for Run 2 are exhibited in Figure S2, denoting the similar superconducting behavior under pressure. 

\begin{figure*}
	\centerline{\includegraphics[scale=0.23]{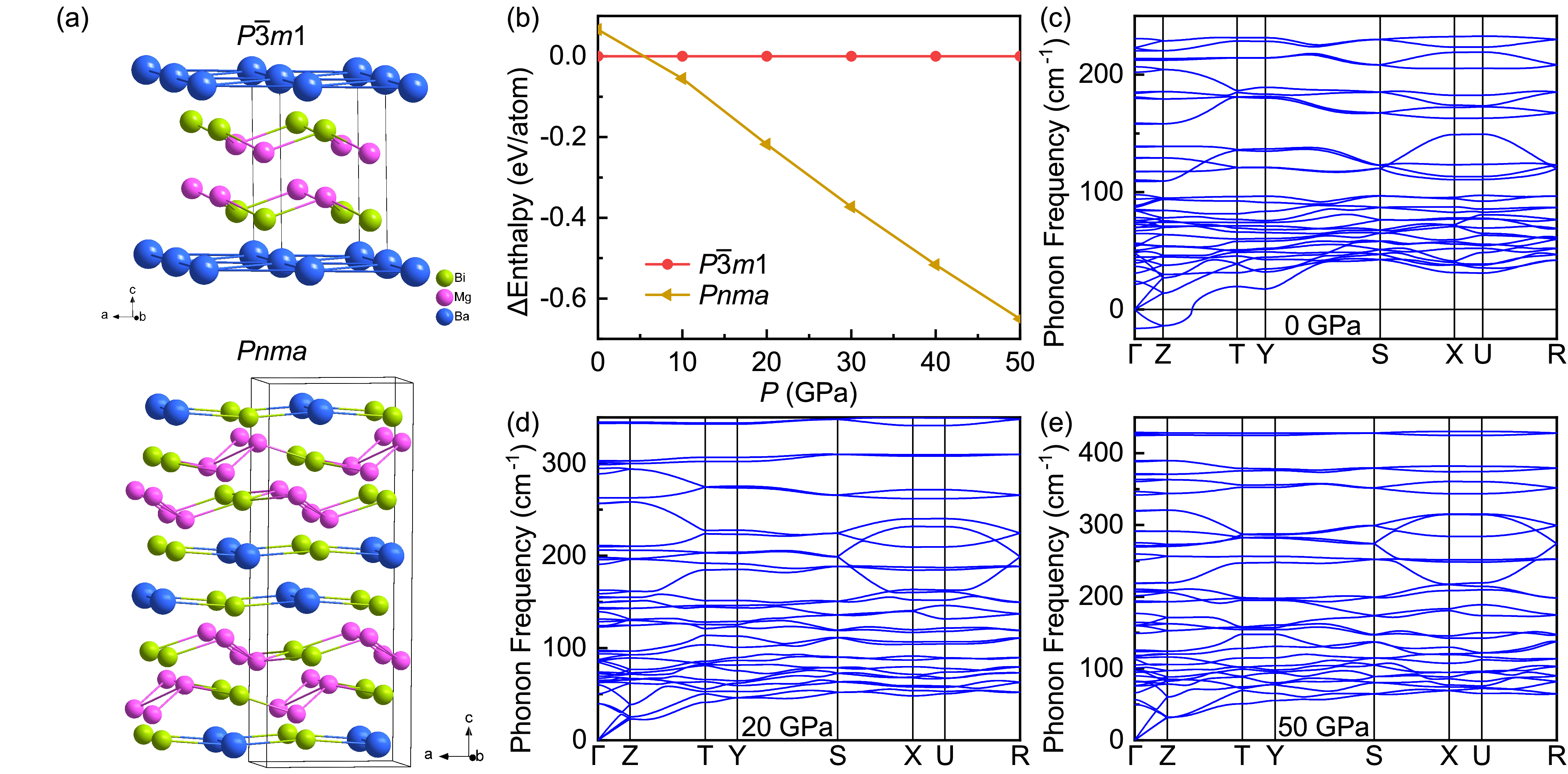}}
	\caption{ (a) The crystal structure of $P \bar{3}m1$ and $Pnma$ phase in BaMg$_{2}$Bi$_{2}$. (b) The enthalpy difference relative to the $P \bar{3}m1$ phase of BaMg$_{2}$Bi$_{2}$ under high pressure. (c)-(e) The phonon spectra of the predicted phase $Pnma$ under different pressures.}
	\label{}
\end{figure*}

\begin{figure*}
	\centerline{\includegraphics[scale=0.25]{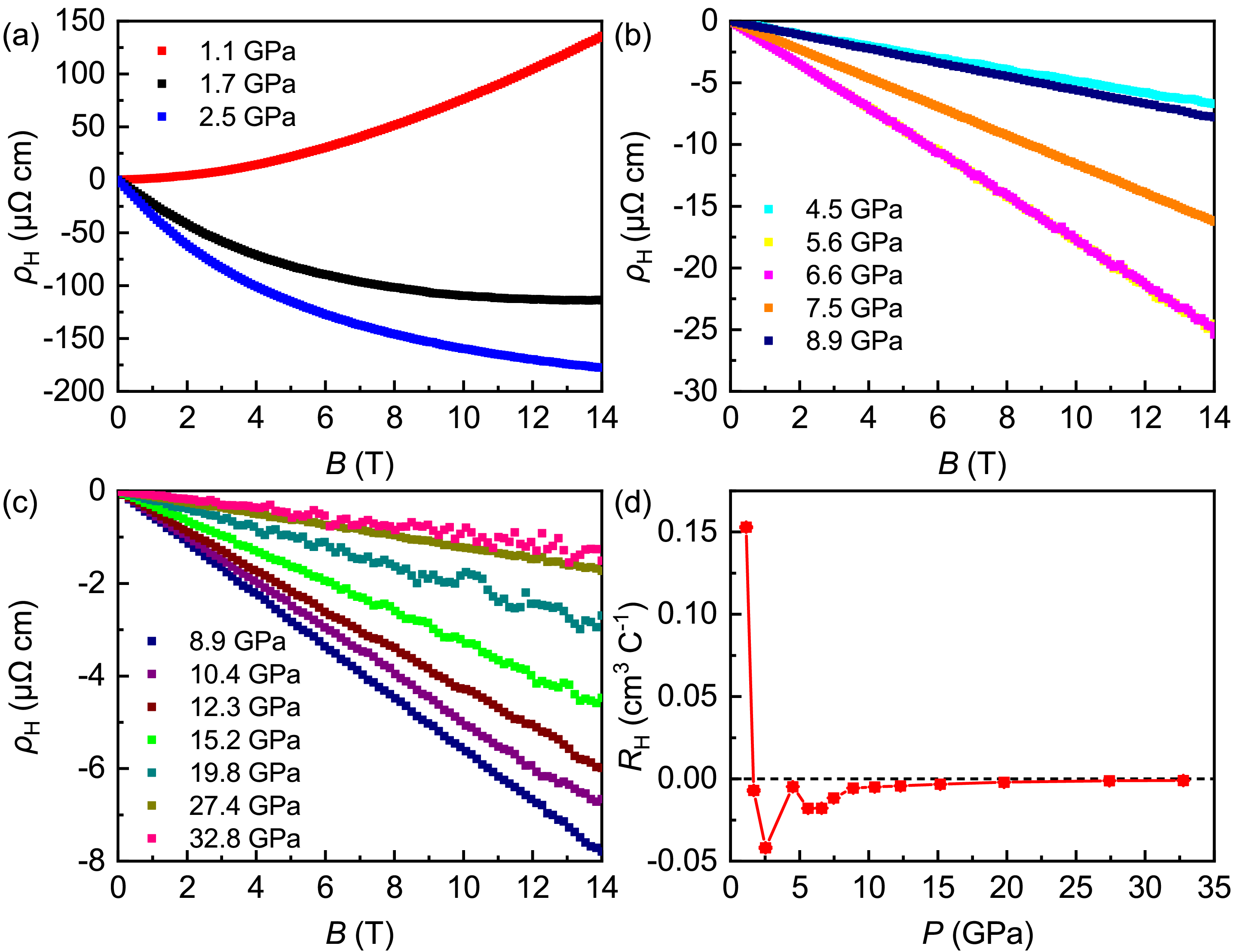}}
	\caption{ (a-c) The Hall resistivity $\rho_{\rm H} (B)$ as a function of magnetic field within the pressure up to 32.8 GPa at 10 K. (d) Pressure dependence of Hall coefficient $R_{\rm H}$. }
	\label{}
\end{figure*}

The temperature dependence of $\rho (T)$ curves under different magnetic fields at representative pressures are shown in Figure 1c-e. $T_{c}$ is progressively suppressed by the application of magnetic field. According to the temperature at which the normal-state resistivity is reduced by 50$\%$, we represent the upper critical field $\mu_{0}$H$_{c2}(T)$ as a function of temperature in Figure 1f. The $\mu_{0}$H$_{c2}(T)$ curve shows negative curvature at the pressure of 1.1 GPa, coinciding with that under ambient pressure and demonstrating the robust surface superconductivity \cite{2208616}. The formula $\mu_{\rm 0} H_{\rm c2}(T) = \mu_{\rm 0} H_{\rm c2}(0)(1-(\frac{T}{T_{\rm c}})^{2})$, where $\mu_{0}H_{c2}(0)$ is the upper critical field at 0 K, is employed to fit the curve. At pressures of 7.5 and 19.8 GPa, the curves turn into exhibiting positive curvature, which can be fitted by utilizing the formula $\mu_{\rm 0} H_{\rm c2}(T) = \mu_{\rm 0} H_{\rm c2}(0)(1-(\frac{T}{T_{\rm c}}))^{1+\alpha}$.\cite{L10} The estimated values of $\mu_{0}H_{c2}(0)$ at pressures of 1.1, 7.5, and 19.8 GPa are 9.24, 7.94, and 1.80 T, respectively. Furthermore, the changes of $\mu_{0}$H$_{c2}(T)$ behavior demonstrate that the surface superconductivity in BaMg$_{2}$Bi$_{2}$ possibly transforms into bulk superconductivity under higher pressures.

\textbf{High-Pressure XRD and Structural Searching}

In view of the pressure-modulated unique superconducting behavior, the evolution of crystal structure in BaMg$_{2}$Bi$_{2}$ under high pressure was investigated. Figure 2a shows the high-pressure XRD patterns of BaMg$_{2}$Bi$_{2}$ single crystal within the pressure up to 25.2 GPa. The Bragg peaks gradually shift to high angles upon compression, which can be indexed by the ambient pressure $P \bar{3}m1$ phase. When the pressure is applied at 10.9 GPa, new diffraction peaks emerge, strongly indicating the appearance of structural phase transition.

\begin{figure*}
	\centerline{\includegraphics[scale=0.6]{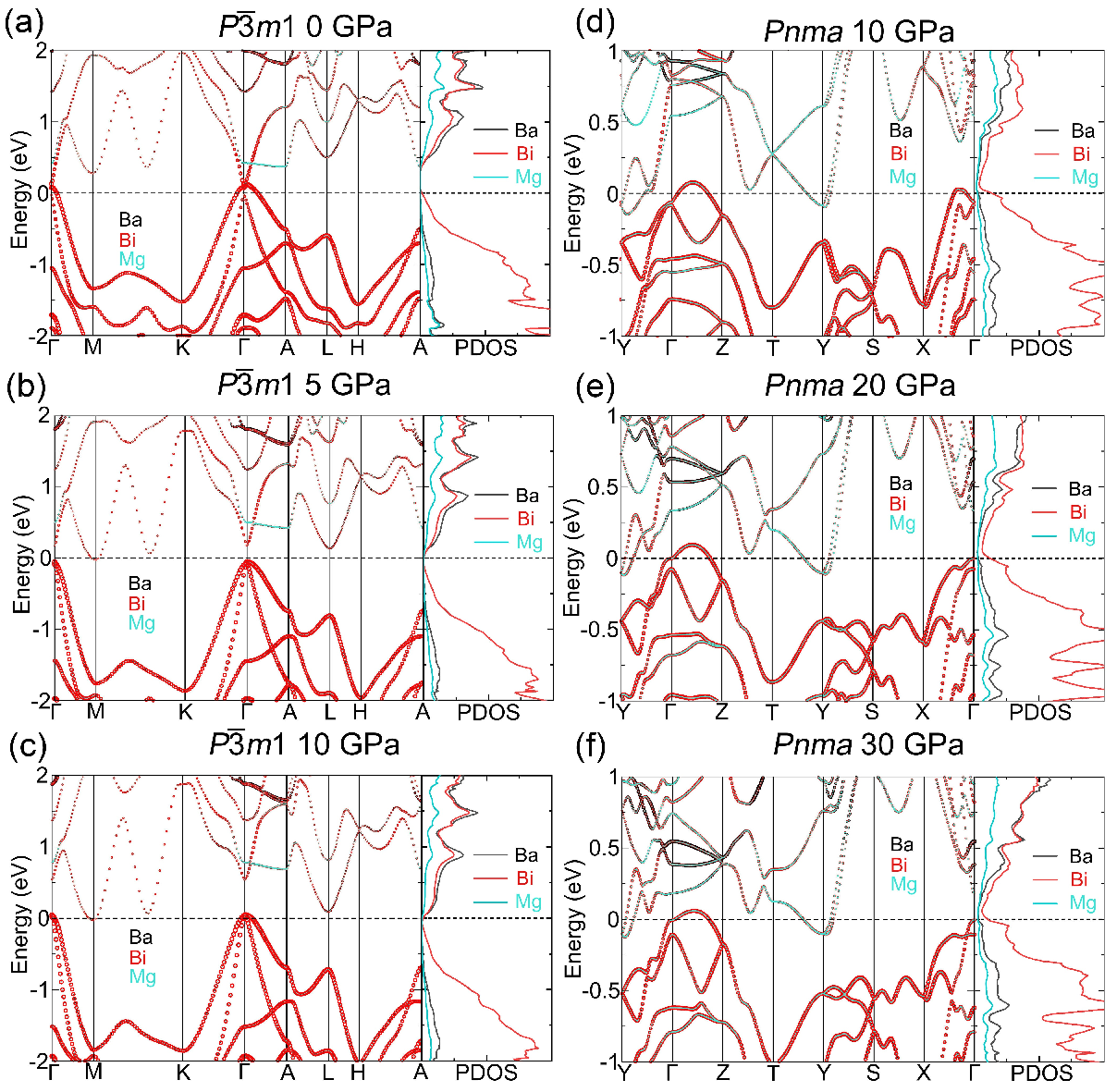}}
	\caption{ The band structures and partial density of states (PDOS) of the $P \bar{3}m1$ phase from 0 to 10 GPa (a)-(c) and the predicted structure $Pnma$ from 10 to 30 GPa (d)-(f).}
	\label{}
\end{figure*}

To identify the crystal structure of BaMg$_{2}$Bi$_{2}$ under high pressure, we performed the high-pressure structural searching. After the structure searches at 20 and 50 GPa in BaMg$_{2}$Bi$_{2}$, we picked out the predicted structure $Pnma$ and compared the enthalpy value with the ambient-pressure structure $P \bar{3}m1$ under high pressure (Figure 3a). As shown in Figure 3b, the enthalpy of the predicted structure $Pnma$ has lower value than that of the $P \bar{3}m1$ phase when the pressure is above $\sim$ 5 GPa, suggesting that the predicted structure $Pnma$ becomes more thermodynamically stable under high pressure. To further check the dynamical stability, we calculated the phonon spectra of the predicted structure $Pnma$ under different pressures in Figure 3c-e. The phonon spectrum of $Pnma$ at 0 GPa has imaginary values about -14 cm$^{-1}$ at $Z$ point (Figure 3c). This illustrates that the predicted structure $Pnma$ is dynamically unstable at ambient condition.
In addition, there are no imaginary values in the phonon spectra of the predicted structure $Pnma$ after 10 GPa (Figure 3d,e), indicating its dynamical stability. Combining the calculations on the enthalpy differences and phonon spectra, the predicted structure $Pnma$ is both thermodynamically stable and dynamically stable above 10 GPa.

According to the structural prediction, it is found that the XRD patterns above 10.9 GPa could be fitted by the predicted $Pnma$ phase (Figure 2b), apparently confirming the occurrence of phase transformation. 
The obtained volume per formula $V/Z$ monotonously decreases with increasing pressure and changes by about 10 $\%$ in the process of structural phase transition, as demonstrated in Figure 2c. 
Furthermore, the $V(P)$ curves could be analyzed by using the third-order Birch-Murnaghan formula 
$P = \frac{3}{2}B_{0}[(\frac{V_{0}}{V})^{\frac{7}{3}}-(\frac{V_{0}}{V})^{\frac{5}{3}}][1+\frac{3}{4}(B_{0}^{'}-4)[(\frac{V_{0}}{V})^{\frac{2}{3}}-1]]$.\cite{135701}  
The fitted volume $V_{0}$, bulk modulus $B_{0}$ for $P \bar{3}m1$ phase is 161.9 $\pm$ 0.5 $\rm \AA ^{3}$, 92.9 $\pm$ 6.7 GPa, respectively; for $Pnma$ phase, the corresponding value is 141.6 $\pm$ 1.7 $\rm \AA ^{3}$, 97.2 $\pm$ 13.3 GPa, respectively.

\begin{figure*}
	\centerline{\includegraphics[scale=0.6]{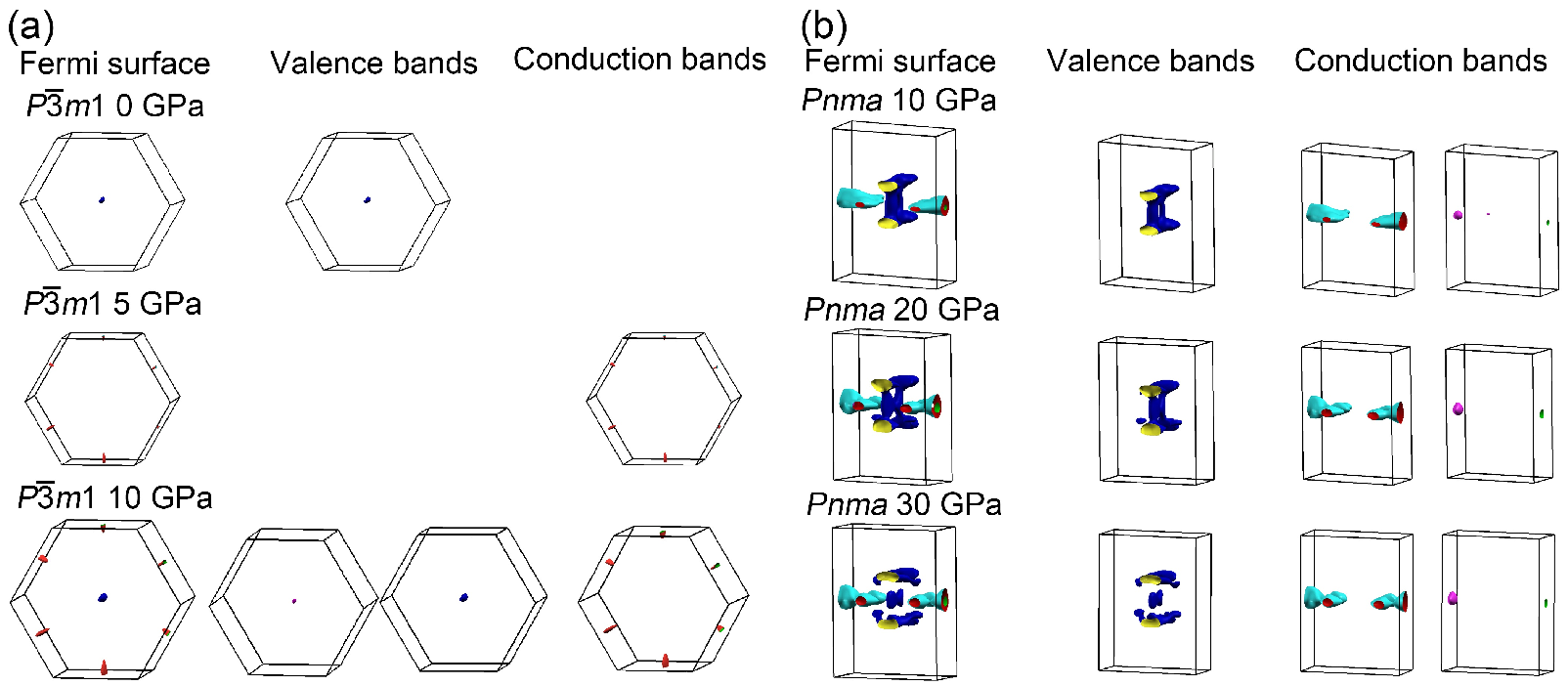}}
	\caption{ The Fermi surface of the $P \bar{3}m1$ phase from 0 to 10 GPa (a) and the predicted $Pnma$ phase from 10 to 30 GPa (b).}
	\label{}
\end{figure*}

\begin{figure*}
	\centerline{\includegraphics[scale=0.38]{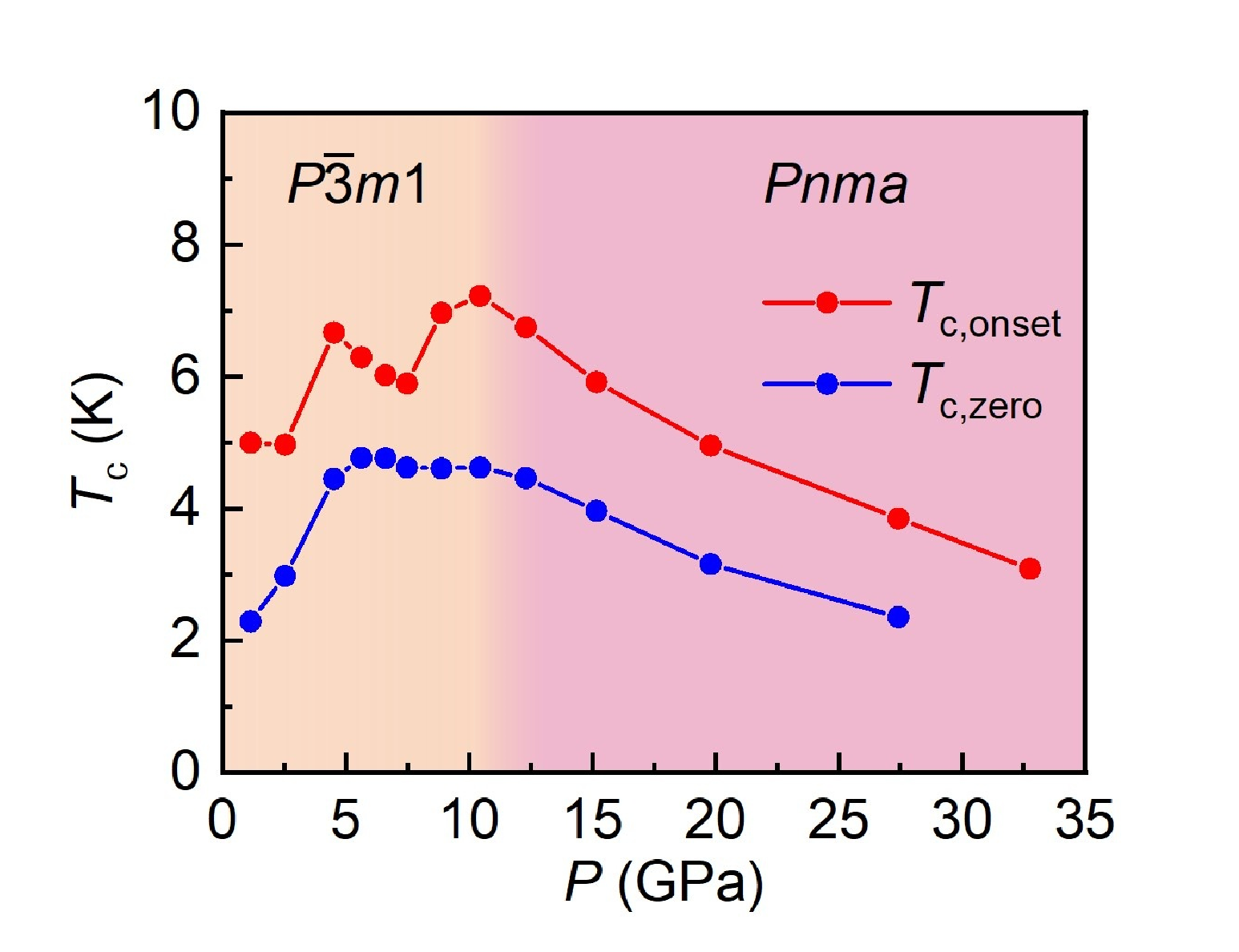}}
	\caption{ The $T_{\rm c}$-$P$ phase diagram of BaMg$_{2}$Bi$_{2}$. }
	\label{}
\end{figure*}

\textbf{High-Pressure Hall Resistivity and Band Structures}

To unveil the evolution of carrier properties under various pressures in the view of electronic transportation, we performed the high-pressure Hall resistivity measurements (Figure 4). The non-linear $\rho_{\rm H}(B)$ curve at 1.1 GPa is positive and demonstrates multiband transport characteristic, as shown in Figure 4a. Nevertheless, at 1.7 and 2.5 GPa, there is a reversal of the sign of $\rho_{\rm H}(B)$ changing from positive to negative. Above 4.5 GPa, the $\rho_{\rm H}(B)$ curves maintain negative and exhibit linear field dependence in the whole magnetic field range, coinciding with the single-band Hall feature (Figure 4b,c). Correspondingly, the Hall coefficient $R_{\rm H}$ under pressures reaching up to 32.8 GPa is illustrated in Figure 4d. Therein, the value of $R_{\rm H}$ below 4.5 GPa was determined by the linear fit of $\rho_{\rm H}(B)$ under high field. It should be noted that the positive $R_{\rm H}$ at 1.1 GPa changes into negative at 1.7 GPa, indicating that the Fermi surface is extremely sensitive to pressure and undergoes a dramatic change. 

To better understand the transport behavior under high pressure, we carried out the calculations on the band structures and partial density of states (PDOS) of both $P \bar{3}m1$ and $Pnma$ phases under high pressure. As plotted in Figure 5, the $p$ electrons of Bi atoms make main contributions to the electronic structures around the Fermi surface in both $P \bar{3}m1$ and $Pnma$ phases. We can observe parabolic crossings around $\Gamma$ point in the $P \bar{3}m1$ phase under ambient pressure, which is in line with the previous reference.\cite{21937,2208616} Besides, valence bands cross the $E_{\rm F}$ (Figure 5a), agreeing with the hole-type carriers in Hall measurements under low pressure. With the pressure increasing (Figure 5b), the parabolic bands around $\Gamma$ point become separate, while the conduction bands around $M$ point cross the $E_{\rm F}$, which shows consistency with the charge carriers type transition to electron-type in Hall resistivity experiments. Although more valence bands cross the $E_{\rm F}$ around $\Gamma$ point after 10 GPa (Figure 5c), the stability calculation indicates that the $Pnma$ phase is more stable. Hence we concentrate on the electronic structures of $Pnma$ phase under higher pressure (Figure 5d-f). The overall electronic structures of $Pnma$ phase persist within 30 GPa, while the variations along the Brillouin zone paths $Y$-$\Gamma$ and $X$-$\Gamma$ around the $E_{\rm F}$ could result in the distinct features on the Fermi surface.  

Thus, we further calculated the Fermi surface of both $P \bar{3}m1$ and $Pnma$ phases under high pressure, as depicted in Figure 6. From 0 to 5 GPa in the $P \bar{3}m1$ phase (Figure 6a), different types of bands cross the $E_{\rm F}$ and cause the vanishing and appearing of the Fermi surface pockets, which corresponds to the charge carriers type transition in the Hall measurements. The transitions of the Fermi surface pockets illustrate the changes of the Fermi surface topology, providing theoretical details of the Lifshitz transition. In Figure 6b, the variations along the Brillouin zone paths $Y$-$\Gamma$ and $X$-$\Gamma$ in the $Pnma$ phase result in the transition from the connected pockets to the isolated pockets in both valence bands and conduction bands, while there are no other pockets emerging or vanishing. Compared to the $P \bar{3}m1$ phase, the relatively conserved changes in $Pnma$ phase could be attributed to the consistent charge carriers type under high pressure. 

\textbf{The $T_{\rm c}$-$P$ Phase Diagram}

The relationship between $T_{\rm c}$ and pressure $P$ for BaMg$_{2}$Bi$_{2}$ single crystal is constructed in Figure 7. An intriguing pressure-driven, M-shaped double superconducting dome is explicitly revealed under high pressure. In the pressure region of the first superconducting dome, where the maximum $T_{\rm c}$ is approximately 6.67 K, there is no structural phase transition that is certified by the high-pressure XRD. By contrast, the types of charge carriers change drastically from being hole-type carrier dominant to electron-type carrier dominant as well as along with the emergence and absence of Fermi surface pockets. It suggests that the first pressure-induced superconducting phase is possibly associated with the Liftshiz transition manipulated by the applied pressure. Moreover, the value of $T_{\rm c}$ is further enhanced when the pressure is above 7.5 GPa and approaches a maximum value of 7.22 K at 10.4 GPa. Combine with the high-pressure XRD and theoretical prediction, the anomalous enhancement of superconductivity is accompanied by the structural transformation from $P \bar{3}m1$ phase to $Pnma$ phase.

\textbf{Conclusion}

In summary, we studied the superconducting properties in Dirac semimetal BaMg$_{2}$Bi$_{2}$ single crystal under high pressure. We present the discovery of pressure-induced double-dome superconductivity, with the maximum $T_{\rm c}$ of about 6.67 K at 4.5 GPa for the first superconducting dome and 7.22 K at 10.4 GPa for second dome. According to the integrated high-pressure XRD, Hall resistivity, and theoretical calculations, it reveals distinct origins for the two superconducting regimes, in which the pressure-tuned Lifshitz transition is closely responsible for the formation of first superconducting phase, and the second one is associated with the pressure-driven structural phase transition. Our study suggests that the Fermi surface topology and lattice symmetry play crucial roles in regulating the exotic superconductivity under high pressure and provides new insights for exploring the unconventional superconducting state in topological materials.

\textbf{Acknowledgments}

This work was supported by the National Key R\&D Program of China (Grants No. 2023YFA1607400), the National Natural Science Foundation of China (No.52272265, 12404161 and 12474098), the Shanghai Sailing Program (Grant No. 23YF1426800). The authors thank the support from Analytical Instrumentation Center ($\sharp$ SPST-AIC10112914), SPST, ShanghaiTech University. The authors thank the staffs from BL15U1 at Shanghai Synchrotron Radiation Facility.

$\dag$ These authors contributed equally to this work.\\

\end{document}